\documentclass[twocolumn,english,amsmath,amssymb,aps,prl,floatfix,nofootinbib,superscriptaddress]{revtex4}
\usepackage[T1]{fontenc}
\usepackage[latin9]{inputenc}
\usepackage{amsmath}
\usepackage{graphicx}
\usepackage{amssymb}

\makeatletter

\providecommand{\tabularnewline}{\\}

\@ifundefined{textcolor}{}
{%
 \definecolor{BLACK}{gray}{0}
 \definecolor{WHITE}{gray}{1}
 \definecolor{RED}{rgb}{1,0,0}
 \definecolor{GREEN}{rgb}{0,1,0}
 \definecolor{BLUE}{rgb}{0,0,1}
 \definecolor{CYAN}{cmyk}{1,0,0,0}
 \definecolor{MAGENTA}{cmyk}{0,1,0,0}
 \definecolor{YELLOW}{cmyk}{0,0,1,0}
 }

\usepackage{graphics}\usepackage{times}\@ifundefined{definecolor}{\@ifundefined{definecolor}{\@ifundefined{definecolor}
 {\usepackage{color}}{}
}{}}{}
\usepackage{bm}
\usepackage{longtable}

\def\be{\begin{equation}}\def\ee{\end{equation}}\def\ba{\begin{eqnarray}}\def\ea{\end{eqnarray}}
\def\la{\langle}\def\ra{\rangle}\def\aB{a_{\mbox{\footnotesize B}}}\makeatother

\usepackage{babel}

\makeatother

\usepackage{babel}

\makeatother

\usepackage{babel}

\begin{document}

\title{Spin Filtering and Entanglement Swapping through Coherent Evolution
of a Single Quantum Dot}

\author{Jose~Garcia~Coello}
\affiliation{Department of Physics and Astronomy, University College London, Gower
Street, London WC1E 6BT, United Kingdom}

\author{Abolfazl~Bayat}
\affiliation{Department of Physics and Astronomy, University College London, Gower
Street, London WC1E 6BT, United Kingdom}

\author{Sougato~Bose}
\affiliation{Department of Physics and Astronomy, University College London, Gower
Street, London WC1E 6BT, United Kingdom}

\author{John~H.~Jefferson}
\affiliation{QinetiQ, Emerging Technologies Group, St Andrews Road, Malvern, Worcs.
WR14 3PS, United Kingdom}

\author{Charles~E.~Creffield}
\affiliation{Departamento de F\'{i}sica de Materiales, Universidad Complutense
de Madrid, E-28040, Madrid, Spain}

\begin{abstract}
We exploit the non-dissipative dynamics of a pair of electrons in
a large square quantum dot to perform singlet-triplet spin measurement
through a single charge detection and show how this may be used for
entanglement swapping and teleportation. The method is also used to
generate the AKLT ground state, a further resource for quantum computation.
We justify, and derive analytic results for, an effective charge-spin
Hamiltonian which is valid over a wide range of parameters and agrees
well with exact numerical results of a realistic effective-mass model.
Our analysis also indicates that the method is robust to choice of
dot-size and initialization errors, as well as decoherence introduced
by the hyperfine interaction.
\end{abstract}

\date{\today}
 \maketitle

{\em Introduction -- } Realizing quantum information and computation
tasks in solid state systems, particularly quantum dots (QDs), has
attracted a lot of interest in recent years. Electron spins in
QDs are promising candidates for the physical implementation of a
qubit \cite{Loss} due to their long coherence times \cite{Taylor}.
Initialization, manipulation, and readout of electron spins have already
been demonstrated \cite{hanson1,petta} and ideas exist for quantum
gates based on single qubits encoded in two QDs \cite{hanson2}. As
it is timely for ``proof of principle'' demonstrations
of multi-qubit processes, it would be highly desirable to establish
a coherent two qubit process in a {\em single} quantum dot.

Bell measurement is a key ingredient that makes possible some important
tasks such as teleportation \cite{teleportation} and entanglement
swapping \cite{entanglement-swapping}. In this Letter, we propose
a mechanism for singlet-triplet measurement based on the {\em coherent}
dynamics of two electrons in a large square QD, followed by a single
charge detection. Such spin-filtering will give a perfect Bell measurement
in the $S_{z}=0$ subspace of two spins. This projection is made possible
due to the existence of a ground manifold of two singlets and two triplets,
separated from higher-lying states by a large energy gap. To a very
good approximation this enables the low-energy coherent dynamics to
be confined to the ground manifold in which the singlets rotate around
the quantum dot whereas the triplets are frozen at their initial locations.
By initializing the system in an unentangled superposition state we
are then able to project onto a singlet or triplet state simply by
a charge measurement to detect whether or not the charge has moved
during the evolution. We use this property to propose some quantum
information applications such as entanglement swapping and generating
the Affleck-Kennedy-Lieb-Tasaki (AKLT) state, which is a resource
for measurement-based quantum computation \cite{Brennen}.

Recently a dissipative method for singlet-triplet measurement has
been implemented experimentally \cite{petta}. In this method a double
QD is prepared with one electron in each QD, and after lowering the
barrier one of the electrons will hop to the other QD provided that
they are in a singlet state. As the singlet state is produced by a
dissipative decay, there is {\em no set time} at which the
electron will hop and the timescale for dissipative relaxation is
usually longer than coherent evolution in the same range of energy.
In our coherent mechanism, however, the operation time is precisely
known and the filtering takes place well within the spin-coherence time.

From a practical perspective a large square QD is easier to fabricate
than a small one and will also be modeled more accurately by our effective
Hamiltonian,
since the energy gap between the
ground manifold and the lowest excited states increases rapidly with dot size,
making the ground manifold increasingly isolated. On the
other hand, as the absolute sizes of the singlet-triplet
splitting in the ground manifold fall exponentially with dot size, large
QDs have slower operation times and are more susceptible to errors.
There is thus a trade-off between these factors, favoring QDs of
intermediate size. Our simulations show that for square QDs of
$L=200-800$ nm our effective Hamiltonian is sufficiently accurate,
and operates at frequencies within the range achieved in recent current experiments \cite{Shinkai}.

{\em Effective Hamiltonian --} We consider a system of two electrons
held in a square semiconductor QD with a hard-wall boundary, which
can be realized in experiment by gating a two-dimensional electron
gas (2DEG) at a heterojunction interface. The spectrum of this system is
determined by the competition between the kinetic energy ($\sim1/L^{2}$)
of the electrons and the Coulomb repulsion ($\sim1/L$) between them.
In small QDs the kinetic term dominates, and the charge density peaks
at the center of the dot (like non-interacting particles). Conversely in large
dots, when the Coulomb interaction dominates, the energy of the system
is minimized by the electrons localizing in space to minimize the
electrostatic interaction energy. In analogy to the concept of the
Wigner crystal state in bulk two-dimensional systems, these highly-correlated
quasi-crystalline states are termed ``Wigner molecules''.

\begin{figure}
\centering \includegraphics[width=0.35 \textwidth]{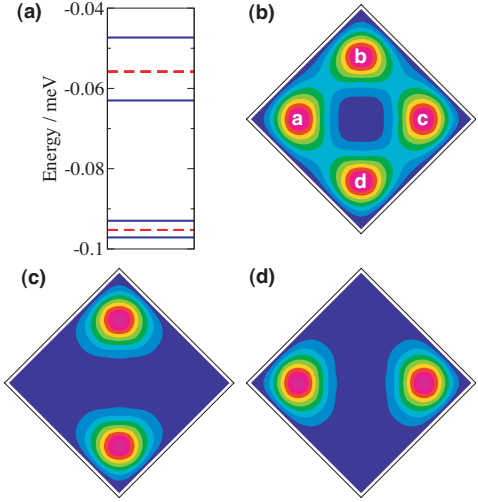}
\caption{(Color online) Eigensystem of a GaAs dot with a side-length of $L=800$
nm, obtained by exact diagonalization of the effective-mass Hamiltonian
(Eq. \ref{effective mass}). (a) The lowest two multiplets of states;
singlets are shown with solid (blue) lines, triplets with dashed (red)
lines. We consider only the dynamics of the lowest multiplet, consisting
of two singlets ($|S_{1}\rangle$ and $|S_{2}\rangle$) with two degenerate
triplets lying between them. (b) Charge distribution of the ground-state,
showing the formation of a Wigner molecule, with peaks labeled $abcd$
near the dot corners. (c) Charge distribution of the symmetrized
singlet state $|1\ra=(|S_{1}\ra+|S_{2}\ra)/\sqrt{2}$, localized about
$bd$. (d) Charge distribution of the antisymmetrised singlet state
$|2\ra$, localized about $ac$.}

\label{Fig1}
\end{figure}

Assuming an effective mass $m^{\ast}$ for the electrons the square
QD is modeled by: \begin{equation}
H=-\frac{\hbar^{2}}{{2m^{\ast}}}\left[\nabla_{1}^{2}+\nabla_{2}^{2}\right]+V({\bf {r}}_{1})+V({\bf {r}}_{2})+\frac{{e^{2}}}{{4\pi\varepsilon|{\bf {r}}_{1}-{\bf {r}}_{2}|}}\label{effective mass}\end{equation}
 where $V(\mathbf{r})$ is the confining potential.
We choose this to be hard-wall with exact square symmetry, though
our results will not qualitatively change under small deviations from
perfect symmetry. The last term in Eq. (\ref{effective mass}) represents
the Coulomb repulsion between the two electrons. In the strongly-correlated
regime, in which the size of the square is large compared with the
effective Bohr radius $\aB$ ($\sim10$nm in GaAs), eigenstates of
this simple Hamiltonian are extremely demanding to obtain exactly.
We show in Fig. \ref{Fig1}(a) the low-lying energy spectrum of a
GaAs QD with side-length 800 nm, obtained by diagonalizing the full
two-electron Schrödinger equation. We see that two degenerate triplets
($|n\ra$, $n=3,4,...,8$) sit approximately (but not precisely) midway
between two singlets ($|S_{1(2)}\ra$), while all these 8 states are
separated from the next multiplet of eigenstates by a relatively large
gap.
The charge distribution for the ground-state $|S_{1}\ra$ is
shown in Fig. \ref{Fig1}(b), and clearly shows how the charge density
strongly peaks near the corners of the QD. One can better appreciate
the form of the states by defining linear combinations of the two
singlets \begin{eqnarray}
|1\ra & = & (|S_{1}\ra+|S_{2}\ra)/\sqrt{2}=|\Phi_{1}^{S}\ra|\psi^{-}\ra\label{States12}\\
|2\ra & = & (|S_{1}\ra-|S_{2}\ra)/\sqrt{2}=|\Phi_{2}^{S}\ra|\psi^{-}\ra,\end{eqnarray}
 where $|\psi^{-}\ra=(|\uparrow\downarrow\ra-|\downarrow\uparrow\ra)/\sqrt{2}$
is the singlet spinor, and $|\Phi_{1(2)}^{S}\ra$ is the symmetric
spatial component of the two-electron wave function. In Fig. \ref{Fig1}(c)
and \ref{Fig1}(d) we plot the charge distribution of these states,
clearly showing how they are localized at diagonally-opposite
corners of the QD. For the triplets we adopt a similar labeling scheme
\begin{eqnarray}
|3\rangle & = & |\Phi_{1}^{A}\ra|\psi^{+}\ra,|4\rangle=|\Phi_{2}^{A}\rangle|\psi^{+}\ra,|5\rangle=|\Phi_{1}^{A}\rangle|\uparrow\uparrow\ra,\label{triplets}\\
|6\rangle & = & |\Phi_{2}^{A}\rangle|\uparrow\uparrow\ra,|7\rangle=|\Phi_{1}^{A}
\rangle|\downarrow\downarrow\ra,|8\rangle=|\Phi_{2}^{A}\rangle|\downarrow\downarrow\ra,\end{eqnarray}
 where $|\psi^{+}\ra=(|\uparrow\downarrow\ra+|\downarrow\uparrow\ra)/\sqrt{2}$,
and $|\Phi_{1}^{A}\ra$ ($|\Phi_{2}^{A}\ra$) is the anti-symmetric
charge distribution, which resembles that of the states $|1\rangle$
and $|2\rangle$, being peaked at the same sites $ac$ ($bd$). Note
that while the triplets $|n\ra$ ($n=3,4,...,8$) are eigenvectors
of $H$, the singlets $|1\ra$ and $|2\ra$ are not.

We can immediately write down an effective Hamiltonian for the low-lying
energy eigenstates \begin{equation}
H_{\mbox{\footnotesize{eff}}} =  -\Delta_{1}|S_{1}\rangle\langle S_{1}|+\Delta_{2}|S_{2}\rangle\langle S_{2}|
 + E_{0}\sum_{n=1}^{8}|n\rangle\langle n|,\end{equation}
 where $E_{0}$ is the energy of the two degenerate triplets, and
$\Delta_{1}$ ($\Delta_{2}$) is the energy separation between the
triplets and $|S_{1}\ra$ ($|S_{2}\ra$). By restricting ourselves
to the ground manifold, and using the sum rule $\sum_{n=1}^{8}|n\rangle\langle n|=I$,
the effective Hamiltonian may be written in the charge-spin form \begin{equation}
H_{\mbox{\footnotesize{eff}}}=E_{0}I-\Delta(|1\ra\la2|+|2\ra\la1|)+J\left({\bf s}_{1}\cdot{\bf s}_{2}-1/4\right),\label{eq:Heff}\end{equation}
 where $J=(\Delta_{2}-\Delta_{1})/2$ and $\Delta=(\Delta_{1}+\Delta_{2})/2$.
This has the simple physical interpretation that the Coulomb
repulsion pushes the electrons to diagonally-opposite corners, giving
two charge states for each combination of spin. Whilst in the corners
the spins of these electrons have an effective Heisenberg exchange
interaction, with exchange constant $J$, and they may tunnel from
one charge state to the other with amplitude $\Delta$.

\begin{figure}
\centering \includegraphics[width=0.35 \textwidth]{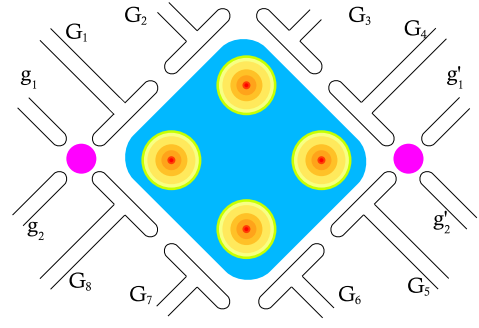}
\caption{(Color online) Gate structure for a large QD (central shaded square),
connected to two smaller QDs (pink circles) at opposite corners.}
\label{Fig2}
\end{figure}

{\em Dynamics --} We now consider the time evolution of two electrons
injected into the square dot so that one is located near
$a$ and the other near $c$ (as labeled in Fig. \ref{Fig1}(b)).
This could be achieved in principle using surface gates as shown schematically
in Fig. \ref{Fig2}. Initially an electron is localized in each of
the small dots adjacent to the large dots. These electrons are then
transferred to the large dot by lowering barriers using gates $G_{1},G_{8}$
and $G_{4},G_{5}$ which are subsequently restored to their previous
potentials after electron transfer has completed. If both electrons
have the same spin, i.e. total $S_{z}=\pm1$, then this spin will
not subsequently change with time under the coherent evolution of
the Hamiltonian (\ref{eq:Heff}) and the two electrons will therefore
remain close to their parent corners. However, if the two injected
electrons are of opposite spin, then the state after injection will
be an equal superposition of a singlet state and an $S_{z}=0$ triplet
state, which {\emph will} subsequently change with time. Let us consider the
state in which a spin-up electron is injected at corner $a$ and a
spin-down electron at corner $c$. We may approximate this state by
\begin{eqnarray}
|\psi(0)\ra & = & \frac{|1\rangle+|3\rangle}{\sqrt{2}}\cr
 & = & \frac{|\Phi_{1}^{S}\rangle+|\Phi_{1}^{A}\rangle}{\sqrt{2}}|\uparrow\downarrow\ra-\frac{|\Phi_{1}^{S}
 \rangle-|\Phi_{1}^{A}\rangle}{\sqrt{2}}|\downarrow\uparrow\ra.\label{eq:initial state}\end{eqnarray}
 Note that both components correspond to spin-up at $a$ and spin
down at $c$ since $\Phi_{1}^{S}+\Phi_{1}^{A}\sim0$ except when ${\bf r}_{1}\sim{\bf r}_{a},{\bf r}_{2}\sim{\bf r}_{c}$
and $\Phi_{1}^{S}-\Phi_{1}^{A}\sim0$ except when ${\bf r}_{1}\sim{\bf r}_{c},{\bf r}_{2}\sim{\bf r}_{a}$.
Hence this state is unentangled.
Under the Hamiltonian (\ref{eq:Heff}), the time-evolution of $|\psi(0)\ra$
can be determined analytically as
\begin{equation}
|\psi(t)\rangle=\frac{e^{-iE_{0}t}}{\sqrt{2}}\left[e^{iJt}\left(\cos(\Delta t)|1\rangle+i\sin(\Delta t)|2\rangle\right)+|3\rangle\right],
\label{evolve}
\end{equation}
 choosing units with $\hbar=1$. We see directly that at time $t^{*}=\pi/2\Delta$,
for which $\sin(t^{*}\Delta)=1$, we have a superposition of the two
states $|2\ra$ and $|3\ra$ with the same probability of finding
either of them. At time $t^{*}$, therefore,
a simple single charge detection at $\emph{any corner}$
(let us say $b$) will project $|\psi(t^{*})\ra$ into a singlet (with
the electrons in corners $b$ and $d$) or a triplet (with electrons
remaining in corners $a$ and $c$). Hence, if we project the state
into a singlet then it oscillates between corners $bd$ and $ac$.
Conversely, if we project it to the triplet then it is frozen in the
corners $a$ and $c$.

The probability of detecting the singlet state at time $t$, starting
in the $S_{z}=0$ subspace, is $P_{2}=|\langle2|\psi(t)\rangle|^{2}=\frac{1}{2}\sin^{2}\Delta t$.
Thus $P_{2}$ oscillates harmonically with maximum probability 1/2
but independently of the exchange, $J$, which simply induces a phase
factor in the singlet component of the wave function. This independence
of $J$ implies that our method of `filtering' the singlet by measurement
is {\em robust} to the size of the dot, for which the ratio $J/\Delta$
falls exponentially with increasing dot size\cite{john1}. This is
not the case for other overlaps. For example, the probability of finding
the initial state is \begin{equation}
P_{\psi(0)}=|\langle\psi(0)|\psi(t)\rangle|^{2}=\frac{1+\cos^{2}\Delta t+2\cos Jt\cos\Delta t}{4}
\end{equation}
 which shows that only for special cases (e.g. $J=0$) does the system
return to its starting state.

\begin{figure}
\centering \includegraphics[width=0.35 \textwidth]{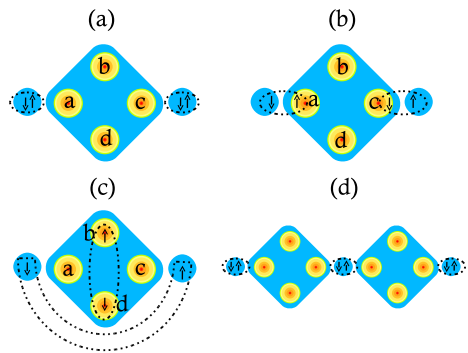}
\caption{(Color online) (a) two small QDs, with a singlet pair in each, beside
a large square QD (dashed lines denote entanglement); (b) One electron
from each singlet is pushed into the square QD; (c) Entanglement swapping;
(d) Scaling up the system to an array of QDs.}
\label{Fig3}
\end{figure}

\textit{Applications --} The ability to make singlet-triplet measurements
paves the way to implement some quantum computation tasks such as
entanglement swapping, or equivalently, teleportation. To achieve
these we generate two singlet pairs outside a square QD as shown in
Fig. \ref{Fig3}(a). These pairs may be generated via surface gates
in a similar fashion to those shown in Fig. \ref{Fig2} in which electrons
are transferred from the surrounding 2DEG
reservoir. The singlets are formed simply by cooling the system \cite{petta}.
We then push one electron from each singlet pair into the big square
QD as shown in Fig. \ref{Fig3}(b). We now have two electrons in the
corners $a$ and $c$ in the square QD and after time $t^{*}$ we
measure the charge at one corner. With probability of $1/4$, the
state of the electrons in the square QD collapses to a singlet at
sites $bd$. In this case the two remaining electrons in the small
QDs get entangled as another singlet, as shown in Fig. \ref{Fig3}(c).
This process is called {\em entanglement swapping} (or the {\em
teleportation} of entanglement) and generates entanglement between
distant particles. This scheme can be scaled up through a geometry
shown in Fig. \ref{Fig3}(d) where a series of empty square QDs are
arranged between small QDs containing electron singlet pairs. By pushing
one electron from each small QD to its neighboring square QD, one
makes all small QDs empty except the two which terminate the array,
that each hold one electron. Dynamical singlet-triplet measurement
on all the square QDs generates a singlet between the electrons held
in the terminating small dots when the result of all measurements
is singlet. The probability of having this is $(1/4)^{N}$, where
$N$ is the number of square QDs.

When the result of measurement in Fig. \ref{Fig3}(b) is a triplet,
rather than a singlet, we can generate the so-called AKLT state \cite{AKLT}.
Originally this was introduced as the ground state of the AKLT Hamiltonian
\cite{AKLT}, which models the interaction of a series of spin-1 particles
with two spin-1/2 particles at the boundaries of a chain. The AKLT
ground state can be generated by again starting with a series of spin-1/2
singlets in small QDs but this time, projecting two particles of neighboring
singlets into a triplet to represent their spin-1 nature. This occurs
with probability $3/4$ when the result of the measurement in Fig.
\ref{Fig3}(b) is a triplet. This can also be scaled up with the geometry
shown in Fig. \ref{Fig3}(d), with probability of success is $(3/4)^{N}$that
all square QD states will be in a triplet state. The AKLT state can
be used as resource for ground-code measurement-based quantum computation
\cite{Brennen}.

\begin{table}
\begin{centering}
\begin{tabular}{|l|c|c|c|c|c|}
\hline
\multicolumn{1}{|l|}{L (nm)} & \multicolumn{1}{c|}{$\Delta$ (meV)} & \multicolumn{1}{c|}{$J$ (meV)} & \multicolumn{1}{c|}{$|\alpha|^{2}$} & \multicolumn{1}{c|}{$|\beta|^{2}$} & \multicolumn{1}{c|}{$E_{hf}(\mu eV)$}\tabularnewline
\hline
100  & 0.814  & -0.243  & 0.441  & 5.23$\times10^{-2}$  & 1.74 \tabularnewline
\hline
200  & 0.145  & -4.363$\times10^{-2}$  & 0.445  & 3.63$\times10^{-5}$  & 7.76 $\times10^{-1}$\tabularnewline
\hline
400  & 2.11$\times10^{-2}$  & -5.05$\times10^{-3}$  & 0.420  & 1.21$\times10^{-3}$  & 3.88$\times10^{-1}$ \tabularnewline
\hline
800  & 2.08$\times10^{-3}$  & -2.20$\times10^{-4}$  & 0.453  & 2.78$\times10^{-4}$  & 1.94$\times10^{-1}$\tabularnewline
\hline
1600  & 9.34$\times10^{-5}$  & -1.66$\times10^{-6}$  & 0.490  & 6.02$\times10^{-6}$  & 9.69$\times10^{-2}$\tabularnewline
\hline
\end{tabular}
\caption{Physical parameters for a GaAs QD. $|\alpha|^{2}$ and $|\beta|^{2}$
(Eq. (\ref{P2e})) are the projection of the initial state onto the
singlet states $|1\rangle$ and $|2\rangle$ by applying a gating
potential of 0.1 V.}

\par\end{centering}

\centering{}\label{values}
\end{table}

{\em Gate Errors --} The above results for the time-development
of the initial state are exact, requiring only the energy parameters
$J$ and $\Delta$, which can be obtained directly from the eigenenergies
of the ground manifold of the effective-mass Hamiltonian Eq. (\ref{effective mass}).
However, these results are somewhat contrived in that the starting
state lies precisely within the Hilbert space of the ground-manifold,
and therefore remains within this ground manifold under time evolution.
In any realistic situation these conditions will not be met and in
particular the starting state will deviate from the idealized form,
Eq. (\ref{eq:initial state}). It will contain small admixtures of
the other base states in the ground manifold and excited singlet states.
These admixtures will increase with decreasing dot size but should
still give small errors for $L>10\aB$, say. We can derive expressions
for the fidelity starting with a more realistic state, $|\widetilde{\psi}(0)\ra$.
This could be produced, for example, by applying a positive potential
to gates located near the sites $a$ and $c$. In the numerical calculations,
this was modeled by dividing the square dot into four quadrants and
applying a constant positive potential to the two diagonally opposite
quadrants that contain the corners $a$ and $c$. In this scheme setting the gating potential to 0.1 V yields values
for the overlap $\la\widetilde{\psi}(0)|\psi(0)\ra$ of 0.80, 0.940,
and 0.97 for QDs of $L=200$ nm, $800$ nm and $1200$ nm respectively,
which are reasonably close to unity, and could be enhanced further
by using more elaborate gating potentials. We may derive an expression
for the fidelity with this more realistic initial state by expanding
$|\widetilde{\psi}(0)\ra$ in terms of $|\psi(0)\ra$, $(|1\rangle-|3\rangle)\sqrt{2}$
and the remaining eigenstates of the full effective-mass Hamiltonian.
After time evolution and projection onto $|2\ra$ we obtain \begin{equation}
P_{2}^{e}=|\langle2|\widetilde{\psi}(t)\rangle|^{2}=(\alpha\sin\Delta t)^{2}-2\alpha\beta\sin Jt\sin\Delta t+\beta^{2}\label{P2e}\end{equation}
 where $\alpha=\langle1|\widetilde{\psi}(0)\rangle$ and $\beta=\langle2|\widetilde{\psi}(0)\rangle$.
Note that $P_{2}^{e}$ is $\emph{independent of excited states}$,
and since $|\alpha|^{2}\sim1/2$, $|\beta|^{2}\sim0$, it is robust
to gate errors. This is illustrated in Table I where we see only small
deviations from the ideal $P_{2}$, even for the smallest dot of $L=100$nm,
the main effect being a suppression of the maxima and enhancement
of the minima.

{\em Charge measurement --}
For simplicity we have so far assumed that charge detection may be made on a
timescale much less than the coherent charge evolution time $t^{*}$. Typical values of
$t^{*}$, however, being of the order of nanoseconds for our parameters
(see Table I) are challenging to measure directly in experiment.
For practical implementation, we propose a similar scheme to Ref. \cite{Shinkai}, which is
able to achieve an acceptable time resolution. At the moment of measurement
we restore the quadrant gate-potentials (used previously to initialize the
system) to freeze the dynamics of the electrons. A strong charge measurement
at one of the corners of the QD can then be made to project the state into
a singlet or triplet.

{\em Charge dephasing --} Charge dephasing reduces the coherence between $|1\ra$ and $|2\ra$ in Eq. (\ref{evolve}), but since our measurement projects onto these
states anyway, it does not fundamentally affect our scheme. By damping the sinusoidal oscillations between $|1\ra$ and $|2\ra$, charge dephasing also reduces
$P_2(t^*)=|\la 2|\psi(t^*)\ra|^2$ such that in the extreme case of very strong decoherence it goes to $1/4$. In this case if $|2\ra$ is detected successfully
the scheme is completed as before, giving entanglement swapping. Otherwise, we end up with a superposition of $|1\ra$ and
$|3\ra$, as in the initial state (except for the amplitude of $|1\ra$ being reduced), which again undergoes damped oscillations. By repeating this process one can reliably (with exponential improvement according to number of trials)
discriminate between singlets and triplets in the initial state.
However, due to our fast dynamics this extreme case is very unlikely.
As an example, for $L=400$ nm we have $t^{*}=0.3$ ns, which is safely below the dephasing
time $T_{2}\sim 1-2$ ns in a system with comparable size \cite{Shinkai}.

{\em Hyperfine interaction --}
Another source of decoherence is the hyperfine interaction between electrons and nuclei
\cite{merkulov-paget}.
This can be estimated by replacing the effect
of the nuclei with an effective magnetic field $\overrightarrow{B}$
coupled to the electron spin as $H_{h}=\hbar\gamma_{e}\overrightarrow{B}.\overrightarrow{\sigma}$,
where $\gamma_{e}=g\mu_{B}/\hbar$ and $\overrightarrow{\sigma}=(\sigma_{x},\sigma_{y},\sigma_{z})$
are the Pauli matrices. $\overrightarrow{B}$ has a Gaussian random
distribution with a variance $B_{n}$ \cite{merkulov-paget}. The
major effect of the hyperfine interaction is to mix the spin-subspaces.
Due to the fast evolution, the first maximum
of $P_{2}(t)$ is relatively unaffected for typical energy values
of $E_{hf}=\hbar\gamma_{e}B_{n}$, given in Table I.
As an example, for $L=400$ nm ($t^{*}=0.3$ns), the decoherence time scale of hyperfine interaction is $10.7$ns.

{\em Conclusions -- } We have shown that the dynamics of a pair
of electrons in a large square QD can be used to perform singlet-triplet
spin measurement using just a single charge detection. This is accessible
to current technology and unlike previous schemes, it is fast, deterministic
and coherent. Repeating the singlet-triplet measurement to a chain
enables entanglement swap over a distance and the generation of AKLT
state in a way that would enable proof of principle quantum information
experiments. Furthermore, coherent evolution of the system is considerably
faster than the dephasing time $T_{2}$ imposed by the hyperfine interaction.
Our low-energy analytic description is valid for a wide range of parameters,
particularly for typical experimental values of the QD parameters.

{\em Acknowledgments -- } AB and SB are supported by the
EPSRC. JGC was supported by the QIPIRC. SB acknowledges the Royal
Society and the Wolfson Foundation. CEC was supported by the MICINN
(Spain) through grant FIS-2007-65723, and the Ramón y Cajal Program.
JHJ acknowledges support from the UK Ministry of Defence.

\end{document}